\documentclass[review,12pt,authoryear]{elsarticle}

\usepackage{amssymb}

\usepackage{amsmath}

\usepackage{color}

\journal{Journal of Aerosol Science}

\begin{document}

\begin{frontmatter}

\title{Development of an electrodynamic balance to study single levitated particles exposed to alkali-metal vapor}

\author{Akira Kamada, Borjid Jiyatai, Atsushi Hatakeyama}

\affiliation{organization={Department of Applied Physics and Chemical Engineering, Tokyo University of Agriculture and Technology},
            city={Koganei},
            postcode={184-8588}, 
            state={Tokyo},
            country={Japan}}

\begin{abstract}
Electrodynamic balances (EDBs) have been widely used
to investigate reactions between levitated particles and background gases. In this paper, we report the development of an EDB that
exposes trapped particles to alkali-metal vapor. The apparatus was developed principally to investigate the interactions between such vapor and the paraffin used as a spin anti-relaxation coating for alkali-metal vapor cells by atomic physicists. The trap electrodes of the EDB were installed in a vacuum glass cell. Particles were loaded via laser launching, without venting or contaminating the cell. Alkali-metal vapor was released from a dedicated dispenser. We found changes in the charge-to-mass ratios of trapped particles irradiated with ultraviolet light after exposure to alkali-metal vapor. These results demonstrate the utility of the apparatus.
\end{abstract}


\end{frontmatter}

\section{Introduction}
\label{sec1}
An electrodynamic balance (EDB) is a device that electrically traps charged particles, and is commonly used in aerosol research. An advantage of an EDB is that it can observe in-situ reactions between single levitated particles and gases. Possible complicating factors such as substrate-particle and particle-particle interactions are not in play \citep{Birdsall2018,Kohli2023}.
One example of such research is the trapping of oleic acid particles interacting with ozone (O$_3$) to investigate the heterogeneous reactions of organic aerosols with atmospheric oxidants \citep{Lee2007}. Another example is trapped sulfuric acid (H$_{2}$SO$_{4}$) droplets under sulfur dioxide (SO$_{2}$) vapor to investigate the Venusian atmosphere \citep{Ubukata2025}. However, no current EDB apparatus adequately observes the reactions of trapped particles with highly reactive metal vapors, particularly those of alkali-metals. Such an apparatus would facilitate research on reactions in several specific environments that contain alkali-metal vapors. These include polar mesospheric cloud
(PMC) particles in the Earth's mesosphere \citep{Plane2015}, and dust in the atmospheres of the moon and Mercury
\citep{Abbas2007,Potter1985,Potter1988}.
Investigation of the reactions between alkali-metal vapors released from biomass,
and ash aerosols generated during biomass combustion, is also important because such reactions significantly reduce the thermal efficiency of boilers \citep{Jöller2007,Niu2016}.

We are particularly interested in the reactions between alkali-metal vapor and solid paraffins. These long-chain saturated hydrocarbons (C${_\textrm{n}}$H$_{\textrm{2n+2}}$) are widely used as spin anti-relaxation coatings for alkali-metal vapor cells \citep{Wu2021}.
Such coatings prevent the spin relaxation of gaseous alkali-metal atoms during collisions with the glass walls of the cells \citep{Robinson1958}.
Alkali-metal vapor cells with paraffin-coated inner walls 
are used for spin-based fundamental physics experiments
\citep{Bao2020,Jin2024,Klein2006,Peng2016},
and to improve the performance of atomic magnetometers \citep{Budker2007} and atomic clocks \citep{Bandi2012}. However, the interactions between alkali-metal vapor and solid paraffin are not fully understood.
For example, the mechanism of the so-called ``ripening" process \citep{Alexandrov2002,Sekiguchi2016}—exposure to alkali-metal vapor to enhance the anti-relaxation performance of coatings—remains unknown \citep{Chi2020,Seltzer2010}.
Some reactions must proceed during ripening, but these remain unclear. 

We used a new EDB to study the interactions between solid paraffin and alkali-metal vapor. An EDB can eliminate the substrate effects that typically degrade reproducibility \citep{Kushida2015,Seltzer2010}. Small changes in trapped paraffin particles caused by exposure to alkali-metal vapor can be detected mechanically \citep{Davis1997}. To demonstrate the feasibility of our approach, we built a new EDB that exposed trapped particles to alkali-metal vapor. The EDB features a vacuum chamber that allows the release of highly reactive alkali-metal vapor from a commercial dispenser during trapping of paraffin particles supplied to the EDB without opening the vacuum chamber. Instead, the particles are launched via laser-induced acoustic desorption (LIAD), and one of them is trapped. The paraffin particle is held trapped at $\sim5$ Pa during melting over $1$ hour using a heater that degasses the particle. The particle is then exposed to alkali-metal vapor while trapped. We unexpectedly found changes in the charge-to-mass ratios ($q/m$) of paraffin particles during ultraviolet (UV) light irradiation after exposure to alkali-metal vapor. The results demonstrate that the apparatus can be used to study the interactions between paraffin and such vapors.

The remainder of this paper is organized as follows. In Section 2, we describe the apparatus. In Section 3, we present the experimental results and the discussion, including the details of particle trapping and $q/m$ measurement. Section 4 contains the conclusions.

\section{Experimental apparatus}
\label{sec2}
The apparatus was designed to trap particles in a vacuum, because alkali-metal vapors are highly reactive and therefore easily oxidized. The trap electrodes of the EDB were placed in a vacuum glass cell. This allowed the use of various optical methods for loading and analyzing particles. The methods employed included LIAD to load particles into the trap, laser absorption spectroscopy (LAS) to detect gaseous alkali-metal atoms, and photoelectrical measurement of particle reactions as described below. We used tetracontane (C$_{40}$H$_{82}$, 87087-5G, Sigma-Aldrich, purity: $\geq 95.0\%$, melting point: $80-84^\circ\text{C}$, density: $0.7785$ g/cm$^3$) as the test paraffin and rubidium (Rb) as the alkali-metal vapor. Both are commonly employed by those who study anti-relaxation coatings.

\subsection{The EDB}
\label{subsec2.1}

\begin{figure}[htpb]
\centering
\includegraphics[clip, width=120mm,angle=0]{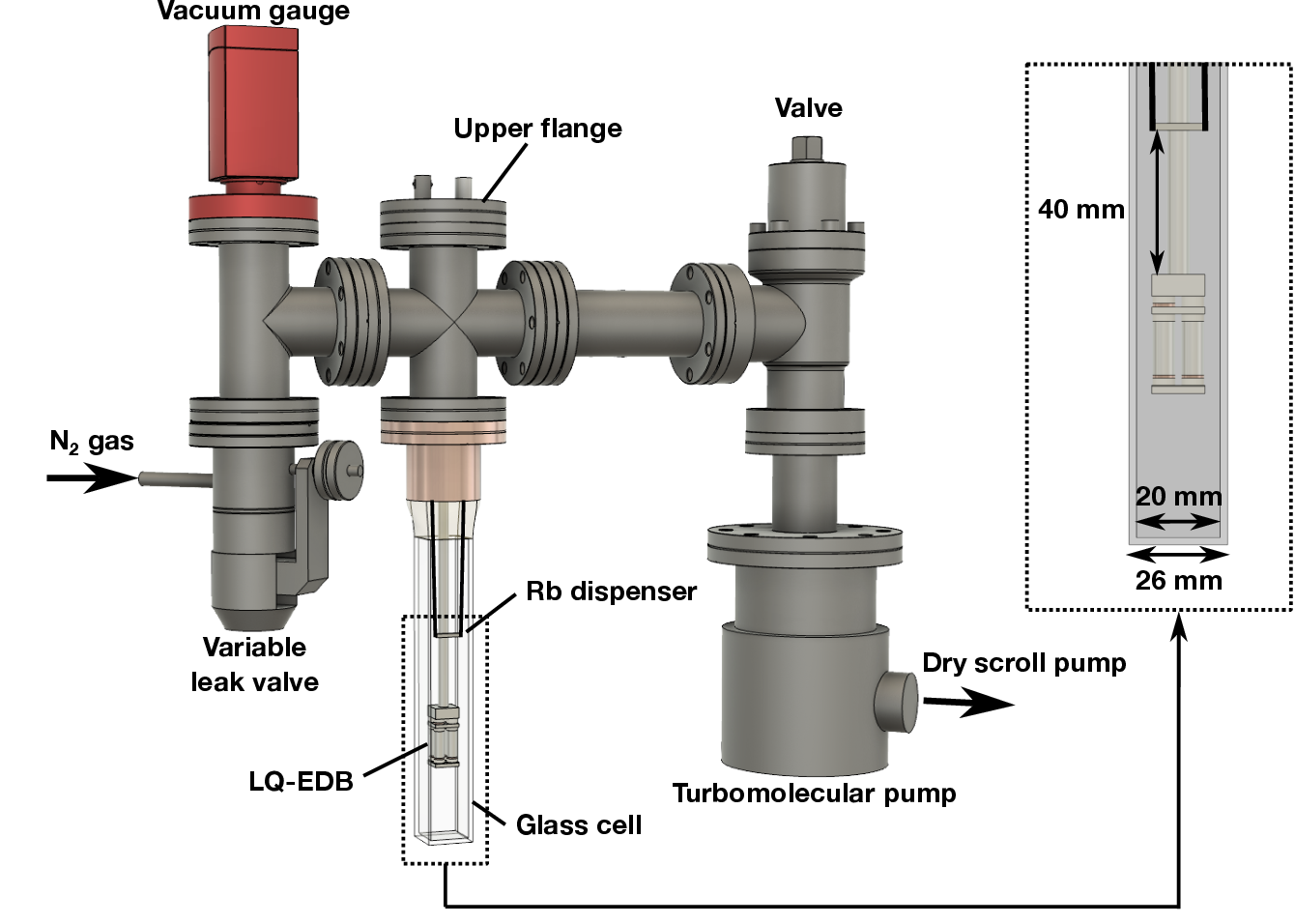}
\caption{
Overall view of the core of the apparatus.
LQ-EDB: linear quadrupole electrodynamic balance.
Inset: magnified image of the glass cell.
}
\label{fig1}
\end{figure}

Figure \ref{fig1} shows an overall view of the core of the apparatus.
The electrodes of the linear quadrupole electrodynamic balance (LQ-EDB) \citep{Hart2015,Walker2024} were installed within the glass cell, as described below. The glass cell was connected to a dry scroll pump and a turbomolecular pump. Nitrogen (N$_2$) gas was used to replace the background gas, thereby creating an inert environment. The LQ-EDB was fixed to the upper ICF70 flange via an aluminum alloy (A5052) rod.
The trapping voltage was applied to the LQ-EDB electrodes by electrical wires via feedthroughs on the upper flange.
The Rb dispenser (RB/NF/3.4/12 FT10+10, SAES Getters), containing Rb chromate (Rb$_2$CrO$_4$) and a zirconium-aluminum (Zr-Al) alloy, was placed 40 mm above the LQ-EDB electrodes and fixed with electrical wires that were connected to the upper flange. On resistive heating at $550-850^\circ\text{C}$ with a current of $4.5-7.5$ A applied to the Rb dispenser, Rb vapor containing $^{85}$Rb and $^{87}$Rb atoms in their natural abundance was released via a reduction reaction, filling the glass cell.
Precise control of the Rb evaporation rate was achieved by adjusting the current applied to the Rb dispenser.
This dispenser was safe to handle in a normal atmosphere because the Rb was stored as a stable compound. Alkali-metal dispensers are widely used in research in material sciences \citep{Lanzoni2022,Xie2020} and atomic physics \citep{Bhardwaj2023,Fortagh2000}. When replacing the Rb dispenser and the tetracontane particle samples, the LQ-EDB and dispenser were removed simultaneously from the vacuum chamber by lifting the upper flange. The method of particle introduction is described below. A ribbon heater attached to the glass cell was used to melt trapped tetracontane particles at temperatures selected by the user. The current supplied to the heater was subjected to feedback control based on the temperature measured by a thermocouple mounted on the exterior of the glass cell. Such particle melting was conducted when degassing trapped tetracontane particles and when initializing the various forms of trapped particles to improve reproducibility.

\begin{figure}[htpb]
\centering
\includegraphics[clip, width=60mm,angle=0]{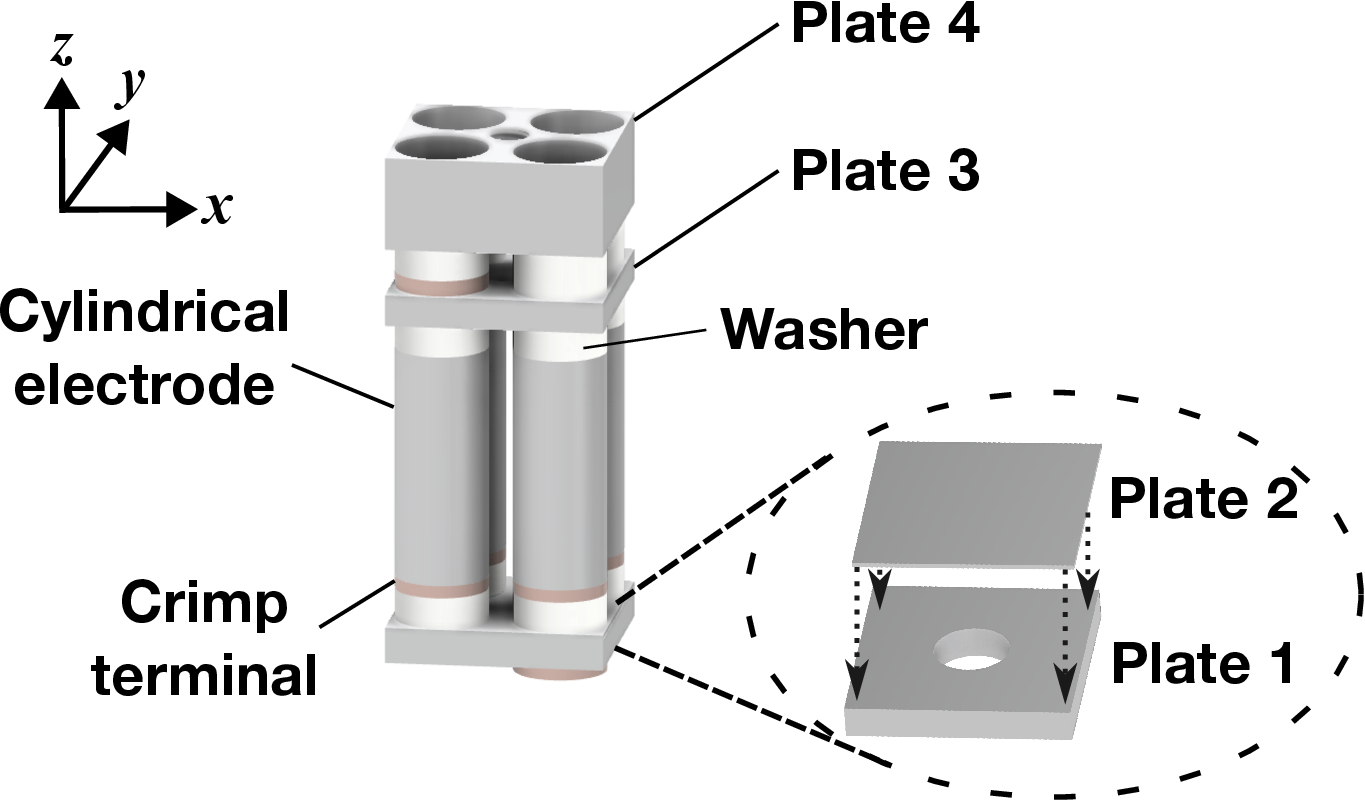}
\caption{Electrodes of the LQ-EDB.}
\label{LQEDB}
\end{figure}

Figure \ref{LQEDB} shows the electrodes of the LQ-EDB. The four cylindrical electrodes confined particles in the $x$-$y$ plane (the horizontal plane). The AC voltage $V_\text{AC}$ with a driven frequency $f_\text{AC}$ was applied to one diagonal pair of the cylindrical electrodes (radius: $3.0$ mm, length: $15.0$ mm), while the other electrode pair was connected to ground. The distance between the trap center and the cylindrical electrode was $2.4$ mm. The cylindrical electrodes featured release holes of diameter $1$ mm. These
were drilled into the screw holes to vent gas trapped during evacuation. The upper and lower planar electrodes confined particles along the $z$-direction (that opposite to gravity). The lower electrode consisted of  plate 1 ($14.0 \times 14.0 \times 2.0$ mm) and plate 2 ($14.0 \times 14.0 \times 0.3$ mm), while the upper planar electrode comprised plate 3 ($14.0 \times 14.0 \times 2.0$ mm). The DC voltage $V_\text{lower}$ was applied to the lower electrode and the DC voltage $V_\text{upper}$ was delivered to the upper electrode. Plate 2 was used by the LIAD to supply particles to the trap, as described below. Plate 4 served as a base from which the LQ-EDB electrodes from the upper flange were hung by the aluminum alloy rod. The cylindrical electrodes and plates 1, 3, and 4 were made of aluminum alloy (A5052), and plate 2 was fabricated from aluminum (A1050), because this was easily thinned. Oxygen-free copper crimp terminals were used to connect the electrical wires to the electrodes. Washers ($96\%$ alumina) served to insulate the electrodes.

\begin{figure}[htpb]
\centering
\includegraphics[clip, width=60mm,angle=0]{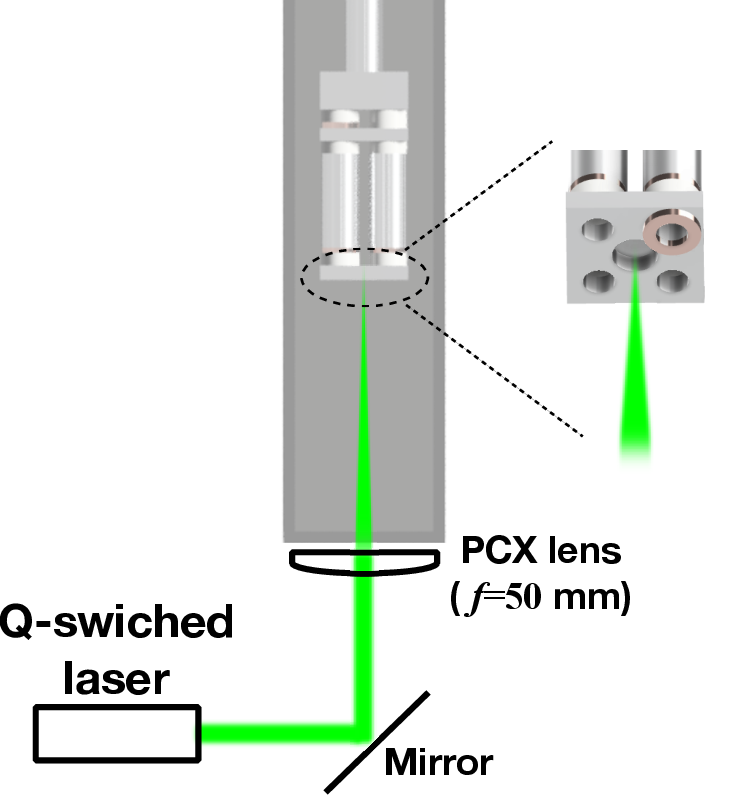}
\caption{The laser-induced acoustic desorption (LIAD) system. PCX lens: plano-convex lens.}
\label{liad}
\end{figure}

LIAD \citep{Asenbaum2013,Bykov2019} was employed as illustrated in Figure \ref{liad}
to introduce particles into the trapping region without venting the vacuum chamber. This minimized the contamination of the chamber associated with
solution spraying \citep{Khorshad2025}.
Tetracontane particles were deposited on the upper surface of plate 2 after removing the EDB from the chamber.
A Q-switched laser (DPS-532-A-5mJ, CNI Laser, wavelength: $532$ nm, pulse width: $8$ ns) delivered one
shot from the bottom as the trapping voltages were applied. Plate 1 featured a central hole of diameter $4.7$ mm through which the laser light was focused onto plate 2. The resulting vibration of plate 2 launched the particles, which were then transported to the trapping region. Launching of particles from the lower electrode plate simplified the apparatus.

\subsection{Charge-to-mass ratio measurement}
\label{subsec2.2}

\begin{figure}[htpb]
\centering
\includegraphics[clip, width=100mm,angle=0]{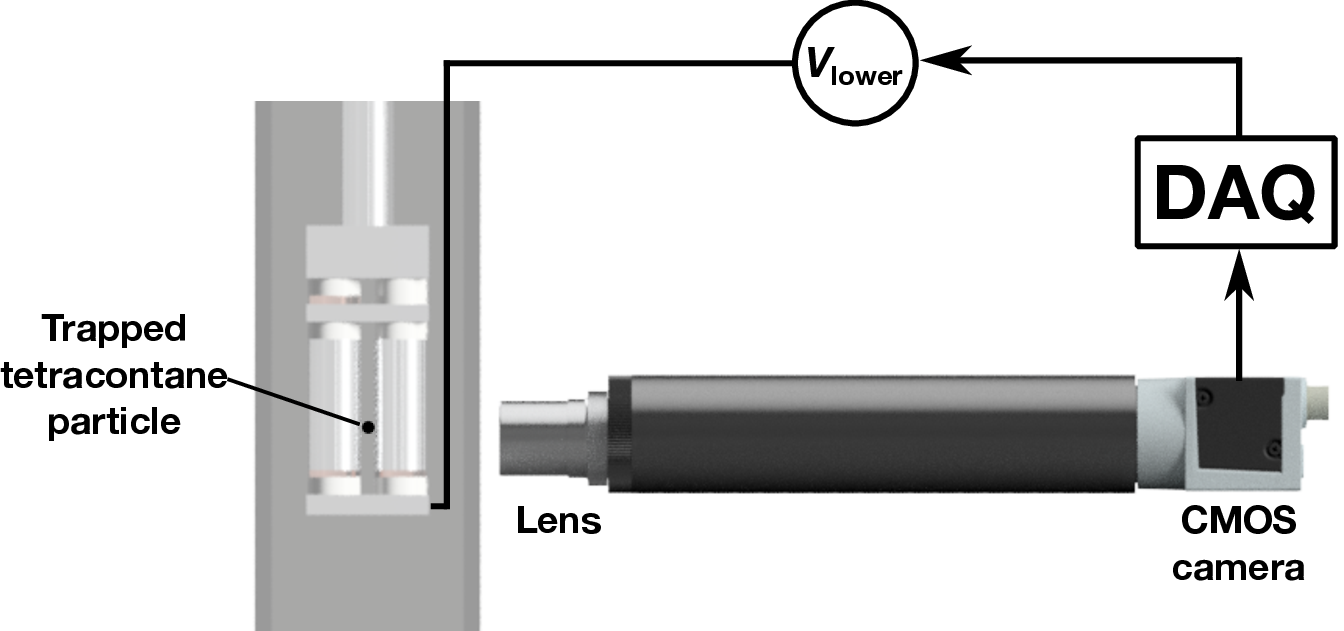}
\caption{
The setup used to measure the charge-to-mass ratio ($q/m$). DAQ: data acquisition system.
$V_\textrm{lower}$: the voltage applied to the lower electrode.
}
\label{q_m}
\end{figure}

The LQ-EDB enables real-time non-contact determination of $q/m$ by monitoring the DC voltage required to maintain a particle at a trapping point. Such measurements have been widely employed to quantify the changes in mass associated with adsorption and chemical reactions \citep{Chen2004}, and to investigate the photoelectric yields of particle surfaces by monitoring variations in the charge of trapping particles \citep{Abbas2007}. In this study, this method was used to detect changes induced by Rb vapor deposition, or chemical reactions, on tetracontane surfaces. The $q/m$ value of a trapped particle was measured using the force balance equation along the $z$ direction between gravity and the Coulomb force exerted by the electric field $E_z$ from the LQ-EDB electrodes:
\begin{equation}
qE_z(z_{\text{particle}})= mg,
\label{qe}
\end{equation}
where $q$, $m$, and $z_{\text{particle}}$ represent the charge, mass, and position along the $z$-axis of the trapped particle, and $g$ is the gravitational acceleration. Figure \ref{q_m} shows the setup used to measure $q/m$. $z_{\text{particle}}$ was derived from images captured
by a complementary metal-oxide-semiconductor (CMOS) camera (acA1300-200uc, Basler). $V_\textrm{lower}$ was controlled to ensure that $z_{\text{particle}}$ was held constant using a PID control of the DAQ system (cRIO-9053 equipped with Linux Real-Time OS and NI-9263, NI) and LabVIEW programs (NI). $E_z$ was evaluated using SIMION 8.2 software (Scientific Instrument Services). The electrode geometries modeled in the software were those of the computer-aided design data shown in Figure \ref{LQEDB}
(grid size: $0.03~\text{mm}$. The washers were modeled as empty). The particle equilibrium position along the $z$-axis was not affected
by the rapidly oscillating electric components originating from $V_\textrm{AC}$. We therefore calculated $E_z$ with $V_\textrm{AC}=0$ V.
Finally, the $q/m$ was determined according to Eq. (\ref{qe}).

\subsection{Rb detection system}
\label{subsec2.3}

\begin{figure}[htpb]
\centering
\includegraphics[clip, width=120mm,angle=0]{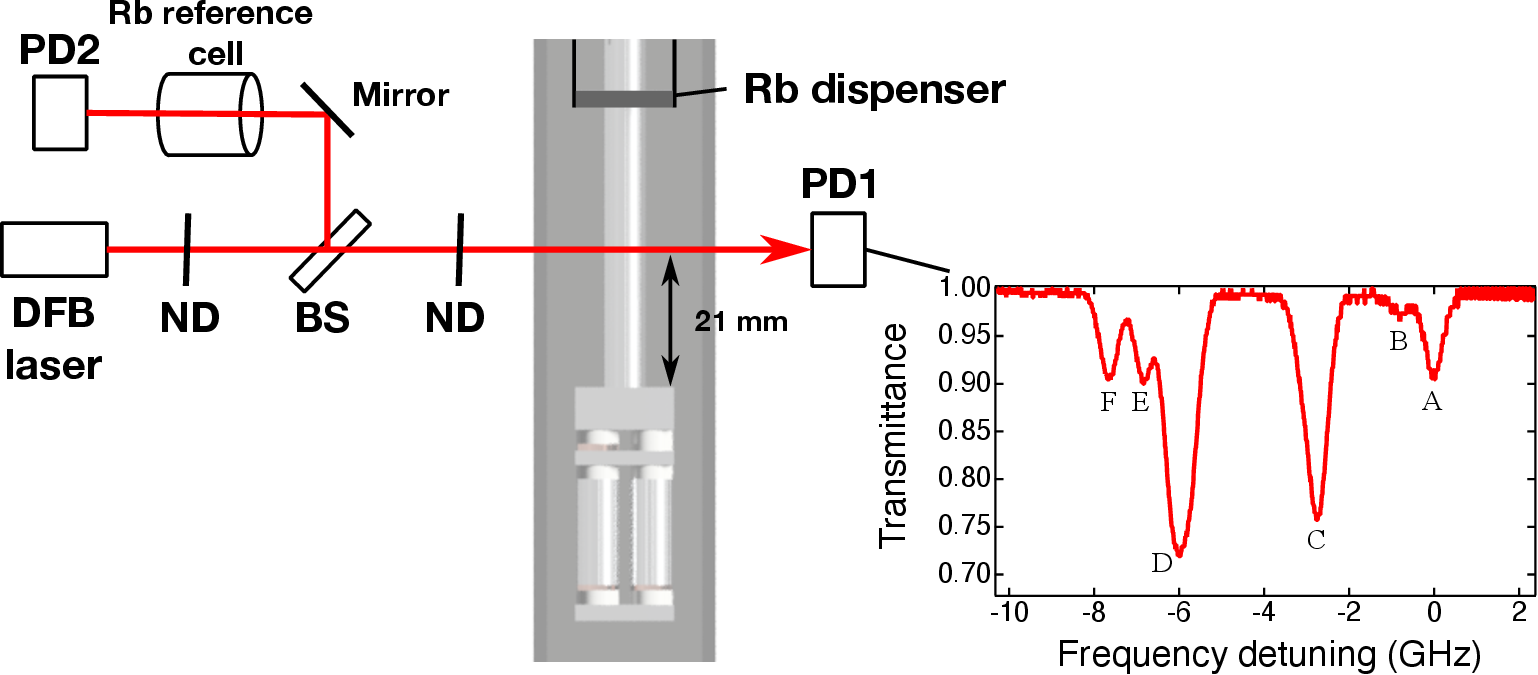}
\caption{
Rb detection system.
DFB laser: distributed feedback laser diode. ND: neutral density filter. BS: beam sampler. PD1: photodetector detecting the laser intensity passing through the glass cell. PD2: photodetector detecting the laser intensity passing through the Rb reference cell. 
Inset graph: a typical absorption spectrum as a function of laser frequency detuning from transition line A.
The transition lines assigned to transmittance dips were as follows \citep{Steck2003,Steck2024,Stern2013}:
A: $^{87}$Rb $F=1\to F'=2$, B: $^{87}$Rb $F=1\to F'=1$, C: $^{85}$Rb $F=2\to F'=2$ and $3$, D: $^{85}$Rb $F=3\to F'=2$ and $3$. E: $^{87}$Rb $F=2\to F'=2$. F: $^{87}$Rb $F=2 \to F'=1$.
}
\label{rb_ab}
\end{figure}

Figure \ref{rb_ab} shows the optical setup for detection of Rb vapor
released from the dispenser and density measurements thereof using the LAS. This allowed highly accurate measurement of Rb vapor density near the trapping site without affecting the electric field of the trap.
A distributed feedback (DFB) laser diode (DFB-0795-0008-BFW01-0005, Eagleyard, wavelength: 795 nm) with current and temperature controllers was used. A laser beam of power $0.6$ $\mu$W was passed through the region between the bottom of the Rb dispenser and the top of the LQ-EDB.  The transmitted laser intensity was detected by a photodetector (PD1). The laser wavelength was scanned across all hyperfine resonance lines of the D1 transitions for both $^{85}$Rb and $^{87}$Rb by changing the laser current. The Rb vapor density in the glass cell was determined from the resulting absorption spectra of the transmitted light \citep{Corney1978}. The scanning wavelength range was calibrated using a reference signal obtained while employing the Rb reference cell.

\section{Results and discussion}
\label{sec3}

\begin{figure}[htpb]
\centering
\includegraphics[clip, width=120mm,angle=0]{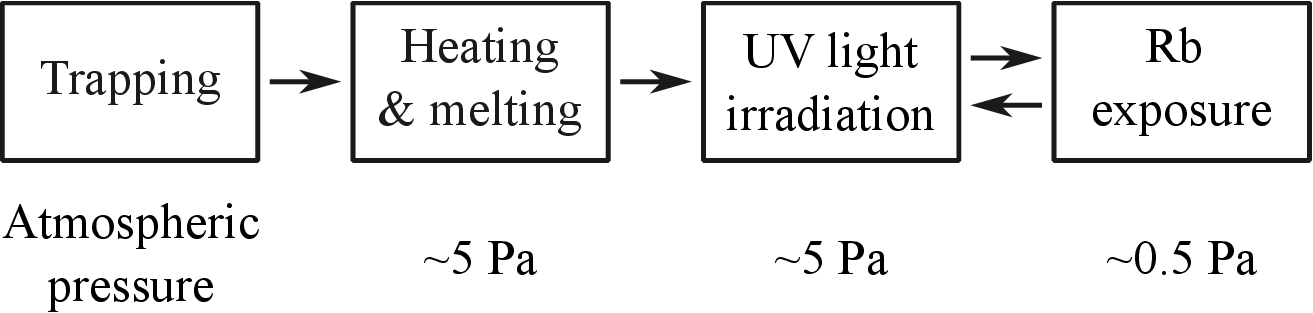}
\caption{Experimental procedure. Typical N$_{2}$ gas pressures in each experimental stage are indicated.}
\label{procedure}
\end{figure}

The experimental procedure is illustrated in Figure \ref{procedure}.
First, a tetracontane particle was trapped.
Next, the trapped particle was heated and melted for degassing.
Finally, the particle was exposed to Rb vapor.
The particle was irradiated with UV light before and after Rb exposure.
Further details are described in the following sections.

\subsection{Trapping}
\label{subsec3.1}
To demonstrate successful launching and trapping, tetracontane particles crushed using a stainless steel needle were deposited on the upper surface of the lower electrode (plate 2).
Most particles were between $1$ and $1000~\mu$m in size, as determined by optical microscopy.
The particles were launched in N$_2$ gas at atmospheric pressure; launching and trapping in a vacuum was not possible given the inadequate viscous resistance. On application of  trapping voltages ($V_{\textrm{AC}}$: $2000$ V, $f_{\textrm{AC}}$: $160$ Hz,  $V_{\textrm{upper}}: 0$ V, $V_{\textrm{lower}}:-400$ V) in N$_2$ gas at atmospheric pressure, tetracontane particles (typical size: $100~\mu$m, typical charge: $10^{-13}$ C) were trapped via single-shot pulsed laser irradiation (2.6 mJ/pulse). 
To roughly determine the particle's charge, the particle volume was estimated by assuming a spherical shape with a radius equal to half of the maximum projected length in the image of particle. The mass  was calculated from the volume and density of tetracontane, and the charge was obtained from the measured charge-to-mass ratio from Eq. (\ref{qe}).
No intentional charging method was applied. 
The particles were possibly charged during crushing through fracture charging and triboelectric charging, or through the interaction with the lower electrode to which a voltage was applied.
The numbers of trapped particles varied for each shot. If multiple particles were trapped, $V_\textrm{lower}$ was decreased to reduce the number of particles to one. Positively charged particles were also trapped by applying positive voltages to the lower electrode. On reducing the trapping voltages $V_{\textrm{AC}}$ to $600$ V and $V_{\textrm{lower}}$ to $-50$ V to prevent vacuum discharge during evacuation, particles remained trapped down to the pressure range ($10^{-2}\sim 1$ Pa) required for Rb vapor release in N$_2$ gas when a turbomolecular pump operates. At these pressures, all particles remained stably trapped when the trapping voltages increased.

\subsection{Heating and melting of trapped tetracontane}
\label{subsec3.2}

\begin{figure}[htpb]
\centering
\includegraphics[clip, width=120mm,angle=0]{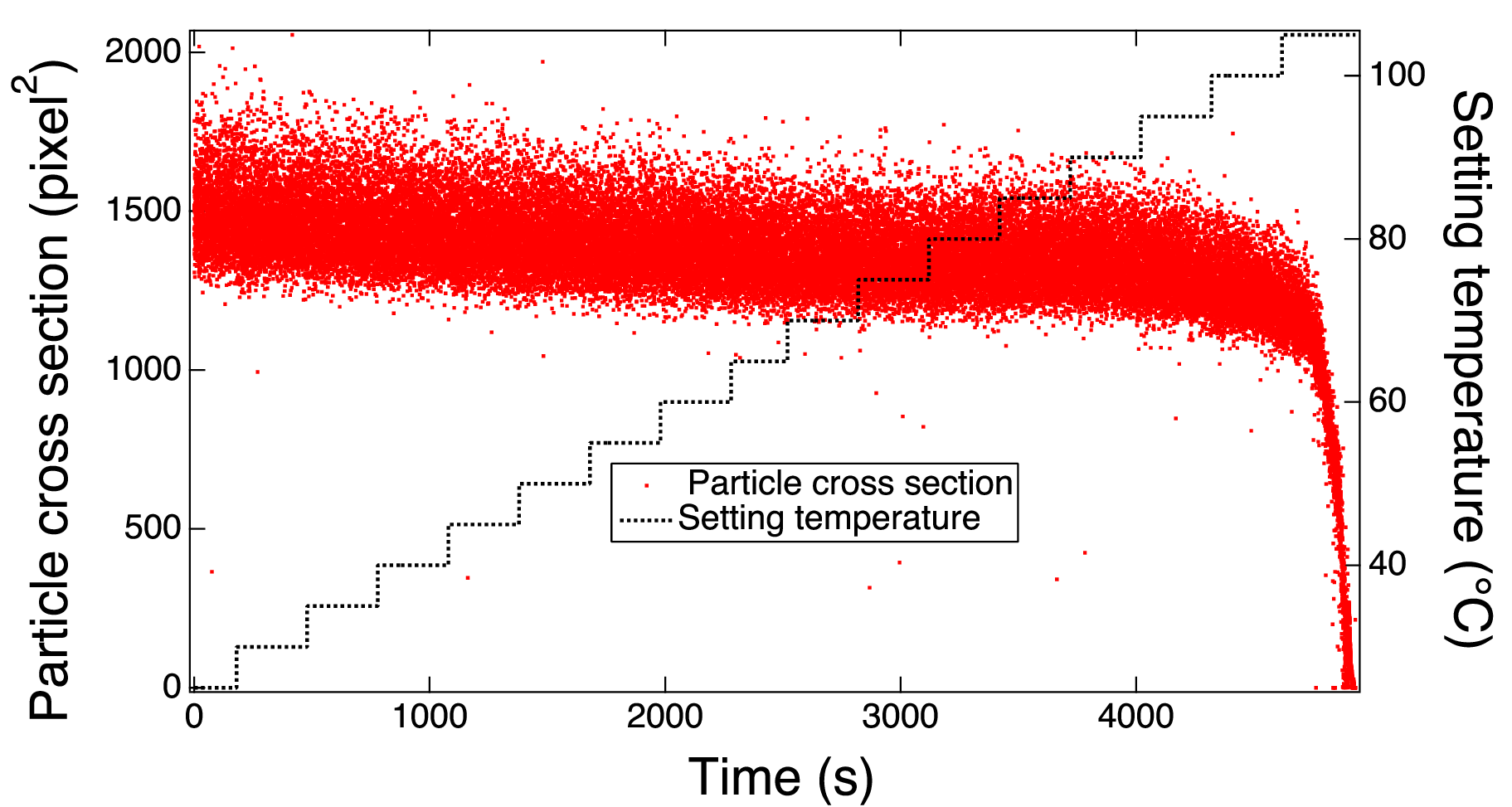}
\caption{Particle cross-section and setting temperature.}
\label{melt}
\end{figure}

A negatively charged tetracontane particle (size: $70~\mu$m) was trapped in N$_2$ gas at $10$ Pa under illumination by a white light-emitting diode (LED) (L-703, HOZAN).
This pressure was set as low as possible to minimize impurities in the particle.
However, it was maintained at this pressure because a further reduction in pressure would destabilize the feedback control of the trapping point.
The setting temperature was increased by $5^\circ\text{C}$ every 5 minutes using the ribbon heater. The measured temperature stabilized within $\pm 2^\circ\text{C}$ approximately 30 seconds after each new temperature was set.

During this procedure, the feedback control of $V_\textrm{lower}$ was activated. Figure \ref{melt} shows the relationship between the setting temperature and the cross-section of the trapped tetracontane particle determined from the number of pixels in captured images that exceeded a certain brightness threshold. This ensured selective detection of the particle. The observed rapid oscillation of the cross-section reflected vibration of both the position and orientation of the trapped particle induced by $V_{\textrm{AC}}$. A significant decrease in the cross-section was observed at $105^\circ\text{C}$. The scattered light fell below the setting threshold while the particle remained trapped. We considered that this reflected melting of the trapped particle because it was contemporaneous with melting of tetracontane particles that remained on the lower electrode, which was observed using another camera (Mx61, SKYBASIC). The reduction in scattered light was presumably attributable to smoothing of the particle surfaces on melting. The discrepancy between the setting temperature ($105^\circ\text{C}$) and the actual melting point of tetracontane ($80-84^\circ\text{C}$) was attributable to the temperature gradient between the external thermocouple and the interior of the glass cell. After resetting the threshold, we observed that the particle was maintained in a melting state for 2 hours at a setting temperature of $107^\circ\text{C}$, and that the cross-section changed only minimally.

\subsection{Rb exposure}
\label{subsec3.3}

\begin{figure}[htpb]  
\centering
\includegraphics[clip, width=120mm,angle=0]{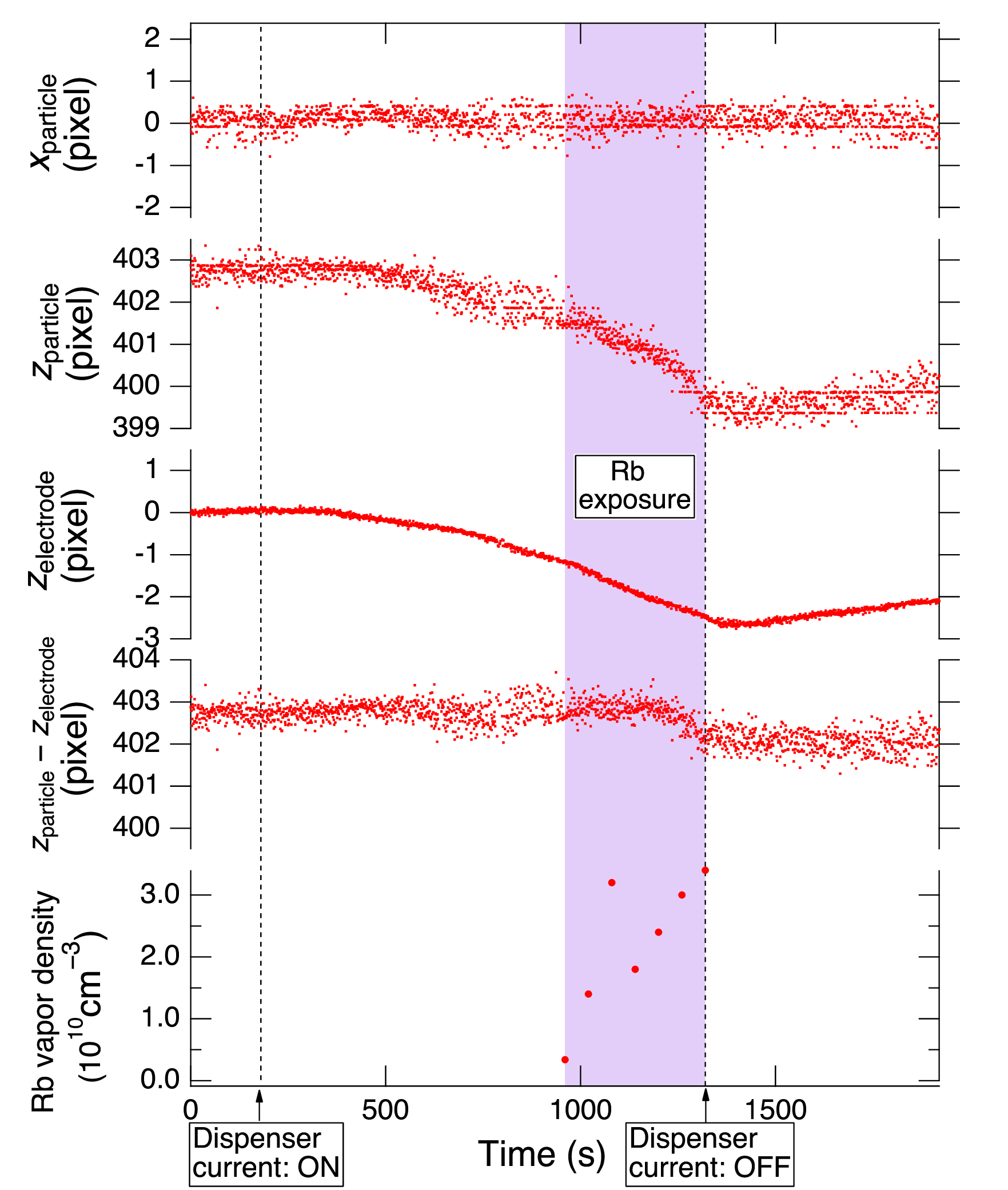}
\caption{
The positions of a trapped tetracontane particle and the top of the lower electrode, the distance between the particle and the lower electrode, and the Rb density when the Rb dispenser was heated.
As the mean free path of Rb atoms was shorter than the typical dimensions of the glass cell in N$_{2}$ gas at $5~$Pa, the Rb vapor density distribution in the glass cell might not be uniform.
The spatial calibration factor was $14.3\pm0.2~\mu$m/pixel.
}
\label{rb_r}
\end{figure}
A trapped negatively charged tetracontane particle (size: $50~\mu$m) was exposed to Rb vapor at $0.5$ Pa in N$_2$ gas.
This pressure was set as low as possible to suppress oxidation of Rb while maintaining stable trapping of the particle. A further reducing pressure decreased the viscous resistance and caused the oscillation of trapped particle to become unstable.
The trapping position was monitored with a camera (Mx61, SKYBASIC) to evaluate trapping stability.
Using the heating system, the particle had been melted at $3$ Pa for $1$ hour while trapped, 16 hours before the measurement. The trapping voltages were set to $V_{\textrm{AC}}$: $600$ V, $f_\textrm{AC}$: $160$ Hz,  $V_{\textrm{upper}}:0$ V, and $V_{\textrm{lower}}: -124$ V. During this measurement, the feedback control of the lower electrode voltage was not activated. To minimize possible light-induced effects during exposure,
a white LED (P6R Core QC, LED LENSER) equipped with a longpass filter (FGL610, Thorlabs, cut-on wavelength: $610$ nm) served as a light source.
All other lights in the laboratory were turned off. Figure \ref{rb_r} shows the $x_{\textrm{particle}}$ (center-of-mass position of the trapped tetracontane particle on the horizontal axis),
the $z_{\textrm{particle}}$ (center-of-mass position of the trapped tetracontane particle on the vertical axis), $z_{\textrm{electrode}}$ (the top of the lower electrode position on  the vertical axis), $z_{\textrm{particle}}-z_{\textrm{electrode}}$, and the Rb vapor density. $x_{\textrm{particle}}$  at $t=0$ was defined as $x = 0$. $z_\textrm{electrode}$ at $t=0$ was defined as $z = 0$. Heating of the Rb dispenser was initiated by increasing the current at 180 s. Rb vapor was first detected at 960 s, and the dispenser current was turned off at 1320 s.
After the dispenser current was turned off, absorption signal of Rb was not detected.
This indicated that the Rb vapor density decreased below the detection limit immediately.
The particle was held trapped during Rb exposure and no significant change was found in the horizontal ($x$) position. This implies that under these experimental conditions the impacts of alkali-metal vapor on the trapping environment—such as discharge and electric field disturbances—were negligible. A downward shift of approximately $40~\mu$m was observed in both the lower electrode
and the tetracontane particle. This displacement was attributable to a downward shift of the LQ-EDB because of thermal expansion of the aluminum alloy rod between the LQ-EDB and the upper flange. However, a small decrease remained in the particle position relative to the 
lower electrode $z_\textrm{particle}-z_\textrm{electrode}$. As this change occurred during Rb vapor exposure, it is possible that an increase in mass or a drop in charge occurred on Rb adsorption or reaction.
However, this change was neither as clear nor as reproducible as that observed when UV light was used to irradiate particles, as discussed below, and requires detailed investigation in future.

\begin{figure}[htpb]
\centering
\includegraphics[clip, width=80mm,angle=0]{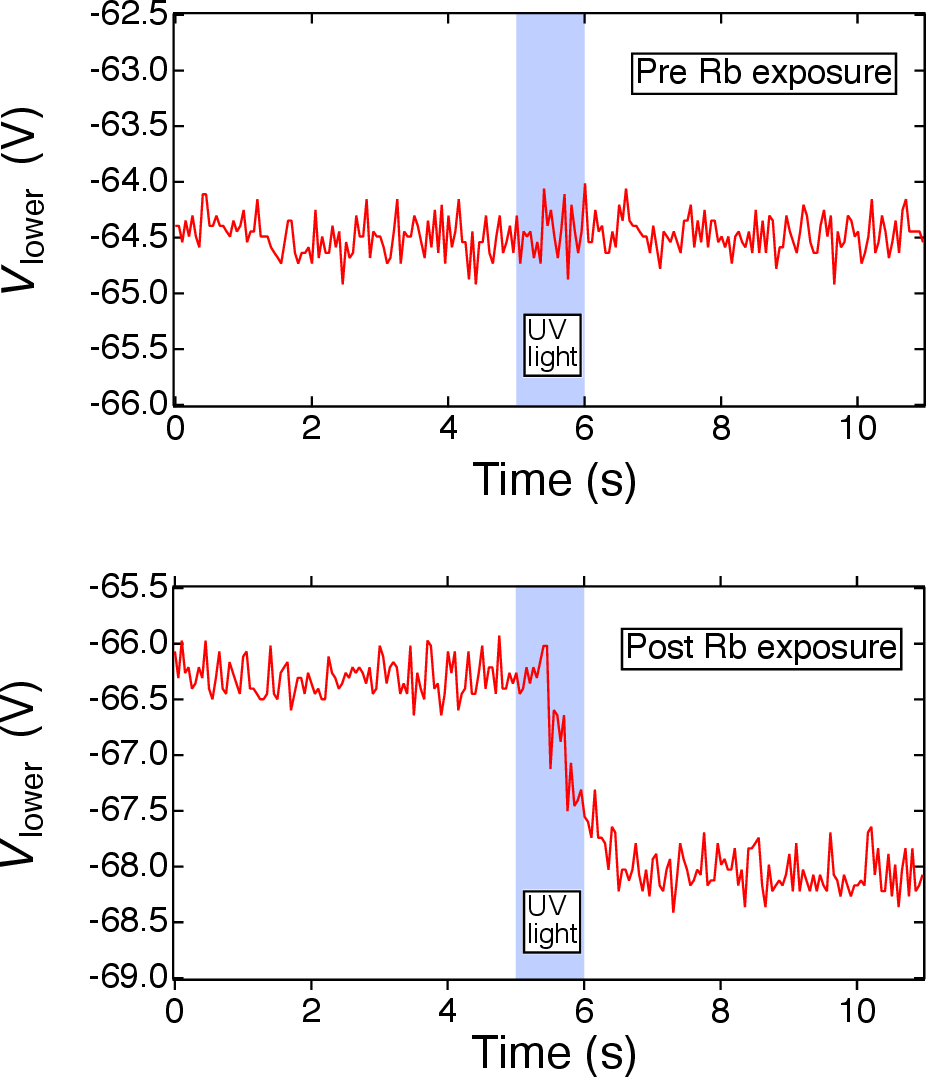}
\caption{Changes in $V_\textrm{lower}$ values under UV illumination for pre-Rb-exposed and post-Rb-exposed particles.}
\label{rb_uv}
\end{figure}

A trapped tetracontane particle was irradiated with UV light at $5$ Pa in N$_2$ gas both before and after Rb exposure (before $0$ s and after $1920$ s in Figure \ref{rb_r}).
This pressure was set as low as possible to suppress oxidation of Rb while avoiding instability in the feedback control of the trapping point.
Figures \ref{rb_uv} show the $V_{\textrm{lower}}$ values under feedback control when the particle was irradiated with a UV laser (NDU7216E, Nichia, wavelength: 374 nm, power: $0.26$ mW, duration: 1 second) before and after exposure to Rb vapor.
Before Rb vapor exposure, no change in $V_\textrm{lower}$ was found during UV light irradiation. However, after Rb exposure, $V_\textrm{lower}$ significantly increased in the negative direction.
For the post-Rb exposure particle, the average $V_\textrm{lower}$ values before and after UV light irradiation were $V_1 = -66.27 \pm 0.03$ V (average over 0.0-1.0 s) and $V_2 = -68.05 \pm 0.05$ V (average over 10.0-11.0 s), respectively. 
Based on SIMION simulations, the electric field was expressed as 
$E_z(z_\textrm{particle}) = \alpha V$, where the factor $\alpha$ was determined
to be $0.028\pm 0.002$ mm$^{-1}$ at the particle position ($z_\textrm{particle}-z_\textrm{electrode}=5.04\pm0.07$ mm).
Employing Eq. (\ref{qe}) and assuming $g = 9.807 \text{ m/s}^2$, the change in $q/m$ was calculated as:

\begin{align}
\Delta \frac{q}{m} &= \frac{g}{\alpha}\left(\frac{1}{V_2}-\frac{1}{V_1}\right)\nonumber\\
&=(1.4\pm 0.1)\times 10^{-4} \text{ C/kg}.
\end{align}
The primary uncertainty was the measurement of $z_\textrm{particle}-z_\textrm{electrode}$.
As no reaction accompanied by mass increase was expected in a N$_2$ atmosphere after termination of Rb vapor release, this change was attributed to a change in charge. 
We repeated the experiment once under identical experimental conditions using a different tetracontane particle, and observed the same reaction.
We further confirmed that this change was not observed when the position of laser irradiation was shifted away from the particle, or when a depleted Rb dispenser (which clearly did not release Rb vapor), was heated. Therefore, we conclude that we detected some interactions between tetracontane particles and Rb vapor with the aid of UV light. The detailed mechanism of the UV-induced effects—possibly involving changes in the surface work function—is currently under investigation.

After venting the glass cell to atmosphere, we found that the surfaces of the electrodes and the glass cell
turned partially white, which was likely caused by the oxidation of deposited Rb.
The electrodes and the cell were therefore cleaned periodically with distilled water and ethanol.
We did not find any effects of deposited Rb on the measurement.

\section{Conclusions}
\label{sec4}
In this study, we developed an experimental apparatus to expose particles trapped in an EDB to alkali-metal vapor. It was specifically sought to investigate the interactions between paraffin and such vapor. The EDB was loaded with tetracontane particles via laser launching, thus without venting/contaminating the vacuum chamber. The trapped particles were exposed to Rb vapor released from an Rb dispenser in vacuum. We confirmed that a solid tetracontane particle was exposed to Rb vapor while still trapped. We detected a clear change in the charge-to-mass ratio of the trapped particle during UV light irradiation after Rb vapor exposure. In future, the apparatus will be used to investigate surface interactions between alkali-metal vapor and solid paraffin serving as a spin anti-relaxation coating. Moreover, the method will aid investigations into the reactions between various alkali-metal vapors released from dispensers and many forms of levitated aerosols.

\section*{Acknowledgements}
This work was supported by JSPS KAKENHI Grant Numbers JP23K26538 and JP24K21729.

\end{document}